\RequirePackage{lineno}
\documentclass[prb,twocolumn,aps,amsmath,amssymb]{revtex4}
\usepackage{graphicx}
\usepackage{bm}
\usepackage{ulem}
\usepackage{color,graphicx,pstricks}
\begin{document}

\title{Impact of Next-Nearest-Neighbor hopping on Ferromagnetism \\ in Diluted Magnetic 
Semiconductors}

\author{Sourav Chakraborty$^1$, Subrat K Das$^{2,}$\footnote{skdiitk@gmail.com}, 
  and Kalpataru Pradhan$^{1,}$\footnote{kalpataru.pradhan@saha.ac.in}}

\affiliation{$^1$CMP Division, Saha Institute of Nuclear Physics, HBNI, Kolkata 700064, India\\
 $^2$SKCG Autonomous College, Paralakhemundi, Odisha 761200, India}

\begin{abstract}

Being a wide bandgap system GaMnN attracted considerable interest after the
discovery of highest reported ferromagnetic transition temperature $T_C$ $\sim$
940 K among all diluted magnetic semiconductors. Later, it became a debate
due to the observation of either very low $T_C$ $\sim$ 8 K or sometimes
absence of ferromagnetism. We address these issues by calculating the
ferromagnetic window, $T_C$ vs $p$, within the $t-t'$  Kondo lattice model
using a spin-fermion Monte-Carlo method on a simple cubic lattice. The
next-nearest-neighbor hopping $t'$ is exploited to tune the degree of
delocalization of the free carriers to show that the carrier localization
(delocalization) significantly widens (shrinks) the ferromagnetic window
with a reduction (enhancement) of the optimum $T_C$ value. We correlate
our results with the experimental findings and explain the ambiguities
in ferromagnetism in GaMnN.

\noindent

\end{abstract}
\vspace{.25cm}

\maketitle
\section{Introduction}

Search for high $T_C$ ferromagnetism in diluted magnetic semiconductors
(DMSs) has been a topic of core importance over the last two decades in view
of potential technological applications~\cite{munekata,ohno1,jungwirth,dietl1}.
A DMS, with complementary properties of semiconductors and ferromagnets,
typically consists of a non-magnetic semiconductor (e.g., GaAs or GaN) doped with
a few percent of transition  metal ions (e.g., Mn) onto their cation
sites. The coupling between electron states of the impurity ions and 
host semiconductors drives the long-range ferromagnetism. The ultimate 
goal is to demonstrate the dual semiconducting and ferromagnetic 
properties of DMS at room temperature.

Mn doped GaAs (GaMnAs)~\cite{ohno,matsukura,okabayashi,ando,dietl,jungwirth1}
is one of the most extensively investigated DMS for which the highest reported
$T_C$ is limited to 200 K [\hspace*{-3px}\citenum{chen,chen1}]. Meanwhile,
wide bandgap based DMSs have attracted substantial attention after the
discovery of room temperature $T_C$ in Mn doped GaN (GaMnN)~\cite{sonoda,reed,thaler}.
Wide bandgap materials are preferred over narrow bandgap semiconductors like 
GaMnAs for two useful reasons: (i) possibility of room temperature
ferromagnetism and (ii) suitability of its band structure for spin 
injection~\cite{kronik}. But, the ferromagnetic state in GaMnN is still a 
debated topic~\cite{sato,dietl2}. In the search of high $T_C$, non-magnetic 
ions (like K, Mg, and Ca) are also considered as potential dopants in
nitride-based wide bandgap semiconductors such as GaN and AlN~\cite{wu,peng,fan,dev1}.
Calculations show that the induced magnetic moment for Ca substitution of 
Ga (single donor) in GaN is 1.00 $\mu_B$~\cite{fan}, while it increases to
2.00 $\mu_B$ for K substitution~\cite{dev1} (K substitution of Ga is a
double donor). Interestingly Ga vacancy induces even a larger magnetic
moment ($\sim$ 3 $\mu_B$)~\cite{fan,dev1,dev} in GaN.

In order to avoid the complication arising due to metal ions,
cation-vacancy-induced intrinsic magnetism are actively investigated in 
wide bandgap nitride-based materials~\cite{dev,larson,missaoui}. The strong
localization of defect states favors spontaneous spin polarization that leads
to the formation of local moments~\cite{dev}. Usually the high formation energy
of such cation vacancies due to unpaired electrons of the anions around the vacancy
sites prohibits us to have enough vacancy concentration that is required for
collective magnetism~\cite{osorio}. Theoretical studies show that the 
formation energy can be reduced by applying an external strain~\cite{kan}.
Overall, there  is  still  no consensus regarding a pathway to engineer 
high $T_C$ nitride-based DMS.

After a considerable amount of efforts has been given to the transition metal
doped GaN based DMS, there is still a lack of fundamental understanding of
the origin of magnetism. In the present work, we focus on certain aspects of
the Mn doped GaAs and GaN like systems using a model Hamiltonian study to address
this fundamental issue. The nature of ferromagnetism in GaMnAs is reasonably well
understood~\cite{jungwirth,dietl1}, and so is regarded as the model system
to understand other similar DMSs. Here, a few percentage of Mn$^{2+}$ ion
(S = 5/2) replaces Ga$^{3+}$, thereby contributing a hole to the host
valence band (VB) which mediate the magnetic interaction between the Mn
spins. But the hole density (holes per Mn ions) is smaller than 1 due to
As antisites~\cite{myers} ($As_{Ga}$) and Mn interstitials~\cite{yu} ($Mn_I$)
which act as double donors. It is well known that co-doping and post-growth
annealing are some effective techniques to alter the hole density~\cite{cho,potashnik}.
These holes reside in the shallow acceptor level introduced by Mn ions in the
host band gap $\sim$ 0.1 eV above the VB~\cite{szczytko,jain,sanvito,sanyal}
reflecting the long-range nature of magnetic interactions between the Mn ions.
If these levels form a {\it distinct} spin-polarized impurity
band (IB) for a finite impurity concentration $x$ then the location of the
Fermi energy $E_F$ will play a crucial role in determining the $T_C$.
Qualitatively, in this IB picture maximum $T_C$ is expected when impurity band
is half filled and supposed to decrease if $E_F$ is near the top or bottom of
the band. In fact, the non-monotonic ferromagnetic window is reported in
experiments for a wide range of hole density~[\hspace*{-3px}\citenum{dobrowolska,chapler}].
This is in agreement (disagreement) with the predictions of Ref.~\hspace*{-3px}\citenum{jungwirth1}
in the high (low) compensation regime, reveling the decisive role of sample
structures along with compensation on the $T_C$ in DMS.

Mn interstitial is the crucial source of compensation as it's removal 
improves both, the hole concentration and the magnetically active Mn ions.
Yu et al [\hspace*{-3px}\citenum{yu1}]
have shown that the $Mn_I$ concentration reduces drastically by diffusing 
from the thin GaMnAs film to the growth surface.
They also showed that all $Mn_I$ can be removed in case of thickness $d$ $<$ 15 nm. 
Due to the effective removal of $Mn_I$ and interfacial effects 
the $T_C$ is reported to be 173 K for $d$ = 50 nm [\hspace*{-3px}\citenum{jungwirth1}],
185 K for $d$ $\sim$ 25 nm [\hspace*{-3px}\citenum{novak,wang}],
and 191 K for $d$ = 10 nm [\hspace*{-3px}\citenum{chen1}].
In comparison to thin films removing $Mn_I$ from the bulk systems
($d$ $\geq$ 60 nm) is difficult, thereby limiting the $T_C$ to 
120 K [\hspace*{-3px}\citenum{ku,dobrowolska,chapler}].
Overall, the hole density, affected by disorder, is very much dependent  
upon the growth process, the thickness and the structure of the DMS. 
In addition, structural defects formed during the growth process can affect 
the electronic structure and hence the $T_C$ of DMS.
In spite of a prolonged and intensive scientific efforts GaMnAs is still
far from the room temperature applications.

GaMnN seems to be a potential candidate to overcome the above issue with $T_C$
over 300 K~[\hspace*{-3px}\citenum{sonoda,reed,thaler}]. However, achieving a
ferromagnetic state in GaMnN is often challenging~\cite{sato,dietl2} and
the physical origin of the ferromagnetism in this material still
remains controversial due to the contradicting experimental
reports~\cite{sarigiannidou,overberg,soo}. In contrast to GaMnAs, Mn is a
deep acceptor in GaMnN forming a distinct narrow IB that is $\sim$ 1.5 eV
above the VB maximum. Consequently, the hole mediated interactions between
the Mn ions are short range in nature~\cite{zunger,graf,janicki,kronik,korotkov,bouzerar}.
Where $p$-type co-doping (Mg in the case of GaMnN) has shown to  enhance
the saturation magnetization~\cite{kim}, the theoretical investigations
found that extrinsic doping of $p$-type generating defects such as Ga
vacancies reduce the stability of the ferromagnetic state~\cite{mahadevan}. In
addition, the coexistence of Mn$^{2+}$ (majority) and Mn$^{3+}$
(minority)~\cite{sonoda1} and the characteristics of defect
states~\cite{mahadevan,sonoda1,sukit} have made the nature of ferromagnetism
in GaMnN more complicated compared to GaMnAs. So the theoretical studies
to understand the ferromagnetism in GaMnN remains elusive to date.

Aim of this paper is to shed light on the unresolved aspects of high $T_C$
ferromagnetism in GaMnN. We consider the $t-t'$ Kondo lattice model and
calculate the magnetic and the transport properties using a
{\it traveling cluster approximation} based spin-fermion Monte-Carlo
method~\cite{tca-ref} on a simple cubic lattice. Degree of delocalization of
the free carriers and hence the magnetic properties are exploited by tuning
the next-nearest-neighbor (NNN) hopping $t'$. We start with a brief
introduction to the model Hamiltonian and the methodology of our approach.
Next, the organization of this paper is threefold: First we establish
appropriate set of parameters for GaMnAs and GaMnN like systems. The
electronic and magnetic properties of GaMnAs are investigated in the
second part. And, finally we calculate and connect our results with GaMnN.

\section{Model Hamiltonian and Methodology}

We consider the diluted Kondo lattice Hamiltonian~\cite{alvarez,chattopadhyay,berciu,pradhan} 

\[H=
-t\sum_{{\langle ij \rangle} \sigma} c^{\dagger}_{i\sigma} c_{j\sigma}  
-t'\sum_{{\langle\langle ij \rangle\rangle} \sigma} c^{\dagger}_{i\sigma} c_{j\sigma}  
+J_H \sum_{m} {\bf S}_{m} .{\vec \sigma}_{m}-\mu\sum_i n_i, \]
\noindent
where $c^{\rm \dagger}_{i\sigma}$ ($c_{i\sigma}$) is the fermion creation
(annihilation) operator at site $i$ with spin $\sigma$. $t$ and $t'$ are the
nearest-neighbor ($\langle ij \rangle$) and the NNN hopping parameters
($\langle\langle$i,j$\rangle\rangle$, respectively. The third term is the Hund's
coupling $J_H$ ($>0$) between the impurity spin ${\bf S}_{m}$ and the itinerant
electrons ${\vec \sigma}_{m}$ (represented by Pauli spin matrices) at randomly
chosen site $m$. We consider the spin $S_m$ to be classical and absorb it's
magnitude 5/2 into $J_H$ without loss of generality. Direct exchange interaction
between the localized spins due to magnetic moment clustering is neglected by
avoiding the nearest neighbor Mn pairing. The overall carrier density $p$
is controlled through the chemical potential ($\mu$) given in the last term. 
$\mu$ is chosen self consistently during the thermalization process to get the 
desired $p$ at each temperature. For impurity concentration $x$ we have
$10{^3}x$ number of spins and $p$ is defined as the holes
per Mn impurity site. We consider $x$ = 0.15-0.25 in a simple cubic lattice,
where as GaAs is face centered cubic with four atoms per unit cell. 
So, the impurity concentration we have taken for qualitative analysis 
in simple cubic lattice is four times to that of fcc lattice. 
Therefore $x$ = 25\% for the impurity concentration corresponds to 
roughly 6.25\% Mn in the fcc systems like GaMnAs~\cite{singh}.
We choose $t$ = 0.5 eV by comparing the bare bandwidth (= 12$t$) of our model to
that of the realistic bandwidth 6 eV for the host III-V semiconductors. Other
parameters such as $J_H$, $t'$, and temperature $T$ are scaled with $t$.

The model Hamiltonian incorporating spatial fluctuations due to randomly
distributed magnetic impurities, as in DMSs, must be carried out for a
reasonably large system size for better results of the physical quantities
such as $T_C$ [\hspace*{-3px}\citenum{alvarez,pradhan}]. We use the exact
diagonalization based classical Monte-Carlo method to anneal the system
towards the ground state at fixed carrier density and temperature. First the
classical spin ${\bf S}_{m}$ is updated  at a site and in this background
of new spin configuration the internal energy is calculated by exact
diagonalization of the carriers. Then the proposed update is accepted or
rejected by using the Metropolis algorithm. A single system sweep composed
of the above processes repeated over each classical spin once. Note that
the exact diagonalization grows as {$\mathcal{O}$($N^4$) per system sweep
and is numerically too expensive for a system size of $N$ = 10$^3$, where 
at each temperature we require at least over 1000 system sweeps to anneal
the system properly. We avoid the size limitation by employing a Monte-Carlo
technique based on travelling cluster approximation~\cite{tca-ref,pradhan1}
in which the computational cost drops to {$\mathcal{O}$($N\times N{_c}{^3}$)
for each system sweep. Here $N_c$ is the size of the moving cluster
reconstructed around the to-be-updated site and the corresponding Hamiltonian
is diagonalized rather than that of the full lattice. This allows us to
handle a system size of $N$ = 10$^3$ using a moving cluster of size
$N_{\rm c}$ = 6$^{\rm 3}$. All physical quantities are averaged over ten
different random configurations of magnetic impurities.

\begin{figure}[t]
\centerline{
\includegraphics[width=9.0cm,height=7.7cm,angle=0,clip=true]{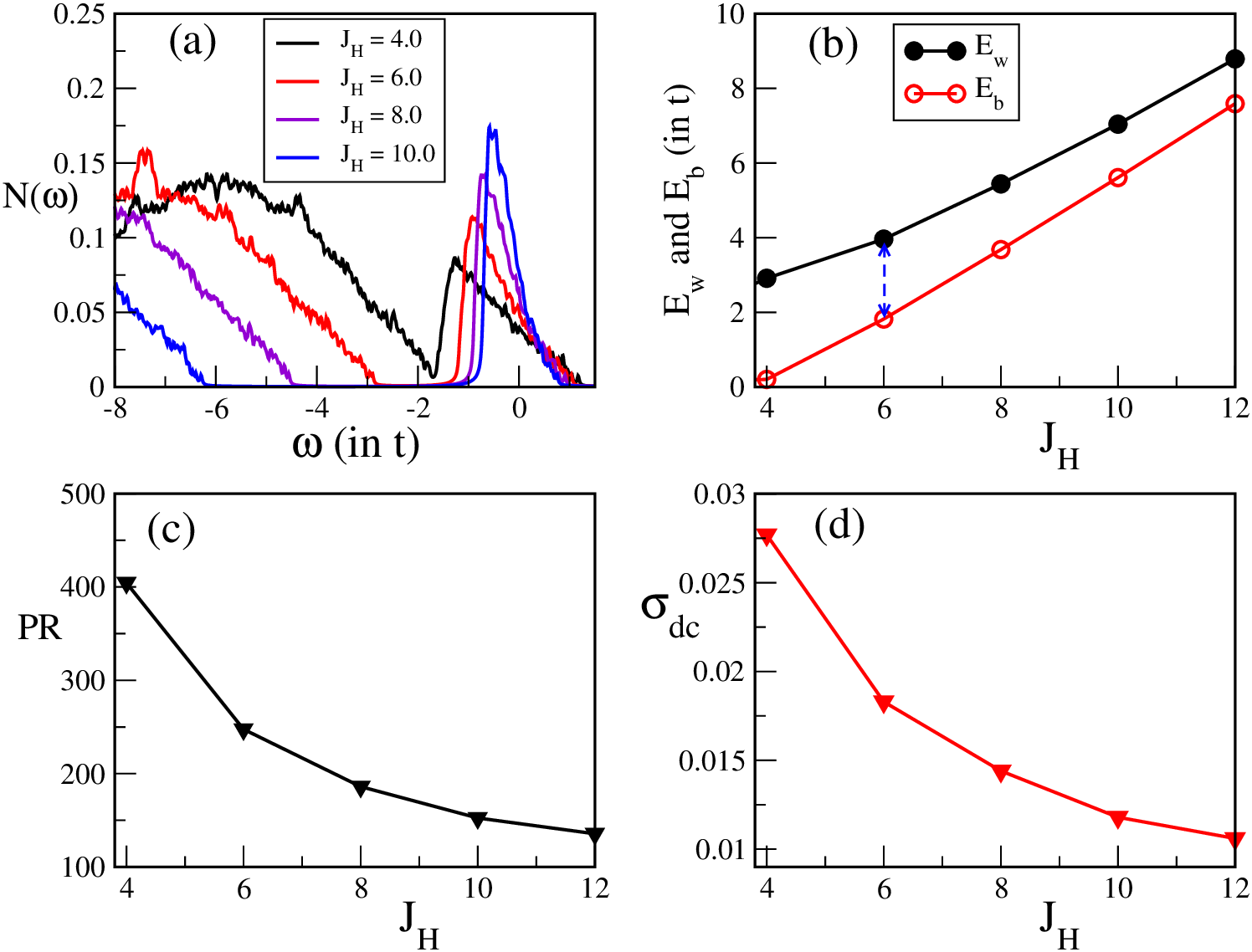}}
\vspace{.2cm}
\caption{\label{fig1}
Shows the (a) density of states $N(\omega)$ with the formation of IB for
different values of the Hund's coupling $J_H$. Fermi energy is set at zero;
(b) change in the binding energy $E_b$ and the $E_w$ with $J_H$ showing
the localization-induced narrowing of the IB (the double arrow shows the
width of the IB for $J_H$ = 6); (c) variation of the participation ratio
with $J_H$ distinguishing the extended states from the localized states, and
(d) decrease in the {\it dc} conductivity (in units of ${\pi  e^2 }/{\hbar a}$)
with $J_H$ indicating carrier localization as in (c). All calculations are
made at fixed impurity concentration $x$ = 0.25, carrier concentration 
$p$ = 0.2, and temperature $T$ = 0.05.
}
\end{figure}

\section{Formation of the impurity band}

The nature of the IB plays the key role in determining the ferromagnetic
state which solely depends on the exchange interaction $J_H$ and the amount
of the magnetic impurities $x$ in the system. Ultra-fast transient
reflectivity spectra~\cite{ishii} and magnetic circular dichroism
measurements~\cite{dobrowolska} show the existence of a preformed IB 
inside the bandgap of in GaMnAs. We start our calculation for $x$ = 0.25,
where a separated IB starts to form for $J_H$ = 4 even at relatively high
temperature $T$ = 0.05 as shown in the density of states (DOS)
$N(\omega)=\langle {1 \over N} \sum_\alpha \delta(\omega-\epsilon_\alpha) \rangle $
in Fig.~\ref{fig1}(a). Here the binding energy
$E_b$ = (bottom of the IB - top of the VB) $\sim$0.2$t$, where the small
finite density of states between the VB and the IB is due to the
broadening used to calculate the DOS. We define the quantity
$E_w$ (= top of the IB - top of the VB), which must be smaller than the
bandgap of the host semiconductor. So $E_w$ - $E_b$ is the width of the
impurity band. With increase in the local Hund's coupling the carriers get 
localized at the impurity sites, consequently the IB becomes narrower and
also moves away from the VB as evident from Fig.~\ref{fig1}(b). From these
results next we fix the $J_H$ values to mimic GaMnAs and GaMnN like systems. 

GaMnAs is a low bandgap ($\sim$ 1.5 eV) system with long-ranged ferromagnetic 
interaction where the $E_b$ is only about $\sim$ 0.1 eV. Hence we choose
$J_H$ = 4 for GaMnAs for which $E_b$ $\sim$ 0.1 eV (0.2t) and $E_w$ is
$\sim$ 1.5 eV (3$t$). Direct measurements yield
$J_H$ = 1.2 eV - 3.3 eV~[\hspace*{-3px}\citenum{okabayashi,ando,matsukura}]
for GaMnAs. Note that we absorbed the impurity spin magnitude 5/2 into $J_H$
which scales with $t$ (= 0.5 eV). So our $J_H$ value is in the range as 
reported in experiments. In contrast, the bandgap of GaMnN is $\sim$ 3.4 eV.
And, the IB is distinctly separated from the VB located at an energy
$\sim$ 1.5 eV ($E_b$) above the VB implying the short-range character of
the ferromagnetic interactions. So in this case we choose $J_H$ = 10, where
$E_b$ $\sim$ 2.75 eV and $E_w$ is $\sim$ 3.5 eV (7$t$). Later, we will see
that the NNN hopping $t'$ hardly alter the $E_w$ value but significantly
affects the ferromagnetic state.

The degree of structural or magnetic disorder is inversely proportional 
to the participation ratio PR $= 1/\sum_i (\psi_l ^{i} )^4$, where $\{\psi_l\}$ 
are the quasiparticle wave functions. PR together with the DOS provide an 
extensive picture of both spectral and spatial features of quasiparticle states. 
The participation ratio provides a measure of the number of lattice sites over 
which the state is extended. For normalized wave functions the PR can 
range between $N$ for a fully extended state and 1 for a site-localized state. 
In Fig.~\ref{fig1}(c) we plot the PR of the state at the Fermi energy
($E_F$) with $J_H$ at fixed $p$ = 0.2 and T = 0.05. For the chosen Hund's
couplings $J_H$ = 4 and 10 the states are extended over $\sim$ 400 sites
and only over $\sim$ 150 sites, respectively. It reflects the fact that
the long- and the short-range nature of the exchange interactions in GaMnAs and
GaMnN like systems are automatically accounted in the calculations.

Then we calculate the {\it dc} conductivity by using the Kubo-Greenwood
formula~\cite{mahan-book, cond-ref} 
\begin{equation}
\sigma ( \omega)
= {A \over N}
\sum_{\alpha, \beta} (n_{\alpha} - n_{\beta})
{ {\vert f_{\alpha \beta} \vert^2} \over {\epsilon_{\beta}
- \epsilon_{\alpha}}}
\delta(\omega - (\epsilon_{\beta} - \epsilon_{\alpha}))
\end{equation}
with $A = {\pi  e^2 }/{{\hbar a}}$, where $a$ is the lattice spacing. 
$n_{\alpha} (Fermi factors) = f(\mu - \epsilon_{\alpha})$ and $\epsilon_{\alpha}$, $\epsilon_{\beta}$
are the corresponding eigen energies. 
And, $f_{\alpha \beta}$=$\langle \psi_{\alpha}\vert$$j_x$$\vert \psi_{\beta}\rangle$
are the matrix elements of the current operator
$j_x = i t  \sum_{i, \sigma} (c^{\dagger}_{{i + x},\sigma}c_{i, \sigma} - h.c.)$.
Finally, the {\it dc} conductivity is obtained by averaging the conductivity over
a small low frequency interval $\Delta \omega$ defined as
\begin{equation}
\sigma_{av}(\Delta \omega)
= {1 \over {\Delta \omega}}\int_0^{\Delta \omega}
\sigma(\omega){\rm d} \omega . \nonumber
\end{equation}
\noindent
$\Delta \omega$ is chosen three to four times larger than the
mean finite-size gap of the system (determined by the ratio of the bare bandwidth
and the total number of eigenvalues). This procedure has been benchmarked
in a previous work~\cite{cond-ref}.
The conductivity for fixed $p$ =0.2 at $T$ = 0.05 is
shown in Fig.~\ref{fig1}(d). The decrease in conductivity with $J_H$
substantiates the fact that the carriers get localized with Hund's coupling 
as seen in Fig.~\ref{fig1}(a)-(c).

\begin{figure}[t]                                                               
\centerline{
\includegraphics[width=9.0cm,height=7.7cm,angle=0,clip=true]{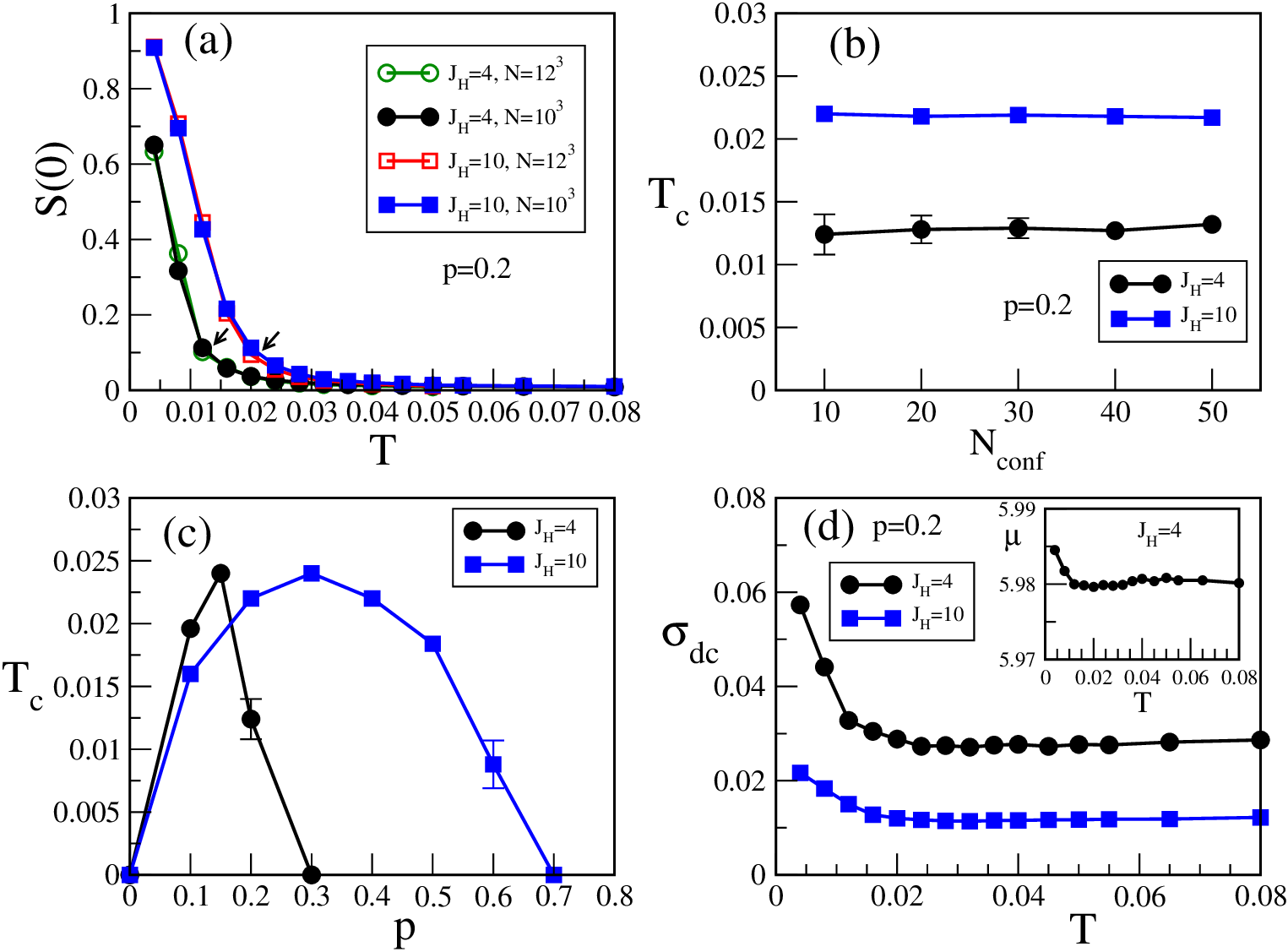}}
\vspace{.2cm}
\caption{\label{fig2}
Displays various physical quantities for $J_H$ = 4 and 10 at fixed $x$=0.25. 
In case of fixed carrier density calculations, $p$=0.2. It demonstrates the
(a) ferromagnetic structure factor S({\bf 0}) for two system sizes 10$^3$ and
12$^3$, which are almost indistinguishable. The arrows point the $T_C$ values;
(b) $T_C$ with error vs the number of configurations $N_{conf}$ clarifying
that $N_{conf}$ = 10 is reasonably good for our qualitative investigations;
(c) ferromagnetic windows $T_C$ vs $p$ showing the localization induced
widening of the FM window in case of $J_H$=10, and (d) {\it dc} conductivity
vs temperature illustrating the more metallicity of the long-range interacting
systems ($J_H$=4) compared to the short-range interacting systems ($J_H$=10).
The inset shows the variation of chemical potential $\mu$ with temperature
for $J_H$=4, required to set the desired $p$ = 0.2.
}
\end{figure}

\section{Ferromagnetic Windows for $t'$ = 0}

In order to see the effects of localization on ferromagnetism we 
estimate the $T_{C}$ from the ferromagnetic structure
factor $S(\textbf{0})$, where $S(\textbf{q})$ =${1 \over N}$ $\sum_{ij}$
$\bf {\bf S}_i\cdot {\bf S}_j$ e$^{i\bf{q} \cdot ({\bf r}_i-{\bf r}_j)}$
({\bf q} are the wave vectors). The average structure factors for $J_H$=4
and 10 are shown in Fig.~\ref{fig2}(a) for $p$=0.2 using system sizes
$10^3$ and $12^3$. As the data of these two system sizes are 
barely distinguishable from each other, so we use $N=10^3$ 
for all calculations in this work. We estimate $T_{C}$ from each structure
factor and then average it over ten different configurations, which is
sufficient enough as shown in Fig.~\ref{fig2}(b). 
$T_{C}$ value remains more or less same with the number of configurations $N_{conf}$.  
The error for $J_H$=4 and $p$=0.2 is found to decrease 
with the number of configurations and is in the acceptable 
range for $N_{conf}$=10 for our qualitative investigations. 
And, for $J_H$=10 and $p$=0.2 the error is insignificant i.e. the error bars are 
smaller than point sizes for all different $N_{conf}$ we considered.

Next we plot the ferromagnetic windows for $J_H$=4 and 10 in 
Fig.~\ref{fig2}(c). The range of the FM window for $J_H$=4 is from $p$=0 
to 0.3. In the higher hole density regime the carriers hopping gets 
restricted due to large delocalization length, and as a result
kinetic energy is minimized and hence the $T_C$ is suppressed. 
On the other hand carriers are less extended for $J_H$=10 
[see Fig.~\ref{fig1}]. Consequently, the carrier hopping is 
stimulated to gain kinetic energy resulting in wider FM window. 
In addition, in Fig.~\ref{fig2}(d) we plot the {\it dc} conductivity 
in a wide range of temperature to corroborate the fact that the carriers 
are relatively more localized for $J_H$=10 as compared to $J_H$=4.
All calculations with temperature are carried out for fixed 
carrier density $p$. The standard procedure to set the desired $p$ at 
all temperatures is by varying the chemical potential $\mu$ accordingly 
with temperature as shown for $J_H$=4 and $p$=0.2 in the inset.

The nature of the carriers that mediate the ferromagnetism and in turn 
controls the $T_C$ depends upon the location of the IB relative to the VB.
Where, for $J_H$ = 4 (GaMnAs-like) the gap is very small, that for
$J_H$ = 10 (GaMnN-like) is large, clearly displaying a separated IB (see
Fig.~\ref{fig1}). Keeping aside the GaMnN case, in literature there are two
conflicting theoretical viewpoints on the nature of the carriers in GaMnAs. 
In one of those extreme limits the IB is very much boardened and indistinguishable 
from valence band, known as the VB picture. In this approach within the mean-field 
Zener model, the magnetic impurities induce itinerant carriers in the VB of
the host materials, which mediate the long-range magnetic interactions~\cite{dietl,jungwirth1}.
It has been generally accepted because of it's ability to explain 
a variety of features of GaMnAs~\cite{dietl,dietl2,dietl3,jungwirth,jungwirth1,
neumaier,nishitani,marco,tesarova,sawicki}. The key prediction of this approach is that 
$T_C$ increases monotonically with both the effective Mn concentration 
and the carrier density~\cite{jungwirth1}. However, this model is contradicted
by electronic structure calculations~\cite{tang,zunger,sanvito1} and argued that
ferromagnetism in GaMnAs is determined by impurity-derived states that are
localized. This is commonly known as the IB picture. Several experiments on the
optical~\cite{hirakawa,burch,sapega,ando1} and transport~\cite{rokhinson,alberi}
properties have reported that $E_F$ exists in the IB within the bandgap of GaMnAs.
Results from resonant tunneling also suggests that the VB remains nearly non-magnetic
in ferromagnetic GaMnAs and does not merge with the IB~\cite{ohya}.  
This picture successfully explains the nonmonotonic variation of $T_C$ with $p$ 
observed in Refs. [\hspace*{-3px}\citenum{dobrowolska,chapler}]. This is in clear
disagreement with the prediction of the valence band picture~\cite{dietl,jungwirth1}.
However, recently it was also suggested that both the mechanisms can be active
simultaneously in GaMnAs~\cite{sato}. In spite of all efforts the issue of IB
picture versus VB picture is still inconclusive.

\begin{figure}[t]
\centerline{
\includegraphics[width=9.0cm,height=4.2cm,angle=0,clip=true]{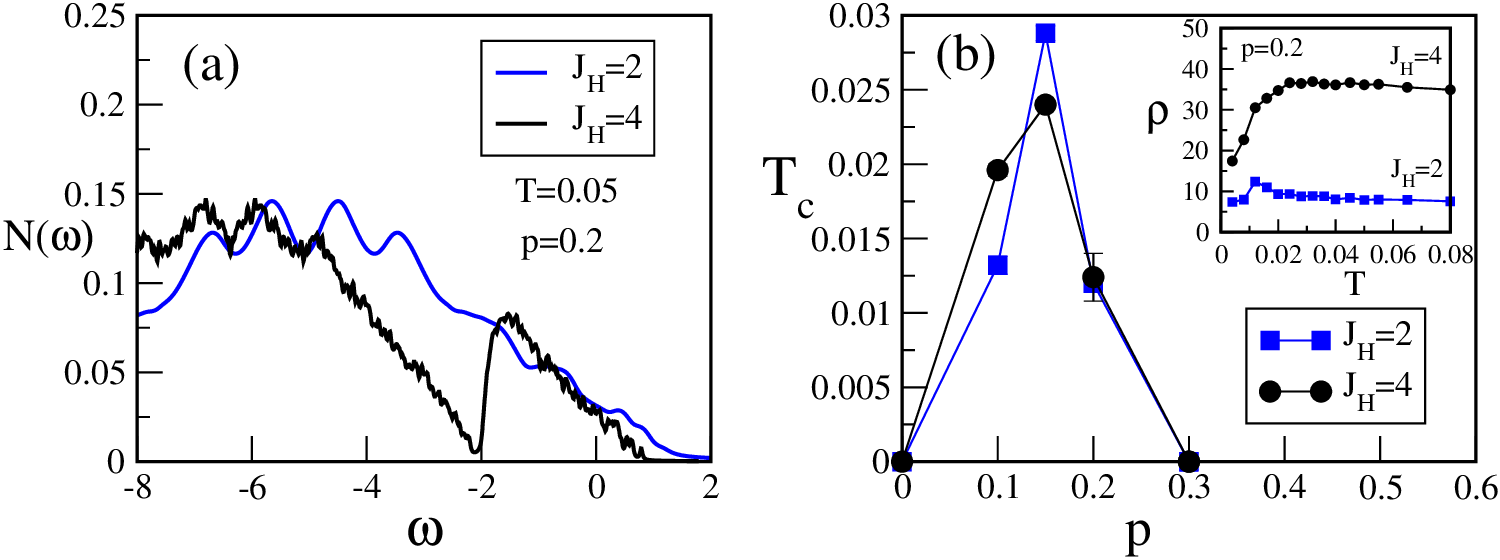}}
\vspace{.2cm}
\caption{\label{fig3}
  Shows the (a) density of states $N(\omega)$ for $J_H$ = 2 and 4 at
  fixed $p$=0.2 and T=0.05. Fermi energy is set at zero. There is no
  signature of IB for $J_H$ = 2,
  and (b) ferromagnetic windows $T_C$ vs $p$ for $J_H$ = 2 and 4.
  Inset plots the resistivity (in units of ${\hbar a}/{\pi  e^2 }$) Vs
  temperature, at $p$=0.2, indicating more metallicity in case of $J_H$ = 2.
  We fixed $x=0.25$ for these calculations.
}
\end{figure}

In this battle of bands~\cite{samarth} where do our assumption of IB picture for 
$J_H$ = 4 in Fig.~\ref{fig1} stands? As we have considered $x$ = $25\%$ in a
N = $10^3$ system, in the ideal situation the IB picture can be assigned
when the participation ratio is within 250 i.e. the carriers are located
only at the magnetic sites. DOS along with PR in Fig.~\ref{fig1} reveal 
that for higher values of $J_H$ (=6 or more) the carriers are restricted to 
the magnetic sites [see Fig.~\ref{fig1}(c)] and so can be categorized in
the IB model. But, in case of $J_H$ = 4 the IB
is very close to the VB and so there is significant probability of hopping
of the holes from the magnetic to host sites. In fact, due to this
hopping process, the participation ratio for $J_H$ = 4 [see Fig.~\ref{fig1}(c)]
is $\sim$400. This shows that there is significant mixing between the VB and IB.
Interestingly, even in the mixed nature of the carriers in case of $J_H$ = 4 the
$T_C$ varies non-monotonically unlike in the valence band picture~\cite{jungwirth1}. 
So there is a natural curiosity to check the $T_C$ trend in the pure VB picture
in our calculation. For this we consider the lower Hund's coupling $J_H$ = 2.
The DOS plotted in Fig.~\ref{fig3}(a) shows that there is no signature of IB at
all. Also, the calculated PR of the Fermi state for $p$ =0.2 is $\sim$800. 
Clearly, this comes in the category of VB picture with more metallicity compared 
to moderately and strongly coupled systems (see the inset of Fig.~\ref{fig3}(b)). 
Most interestingly, the $T_C$ shows an optimization behavior with respect to
$p$ as in the case of $J_H$ = 4 [see Fig.~\ref{fig3}(b)]. So we found that the non-monotonic 
behavior of $T_C$ is independent of the VB and the IB pictures. Similar results
also found by other techniques such as spin wave and earlier MC
calculations~\cite{singh,singh1,alvarez,bui}.

\begin{figure}[t]                                                    
\centerline{
\includegraphics[width=9.0cm,height=7.7cm,angle=0,clip=true]{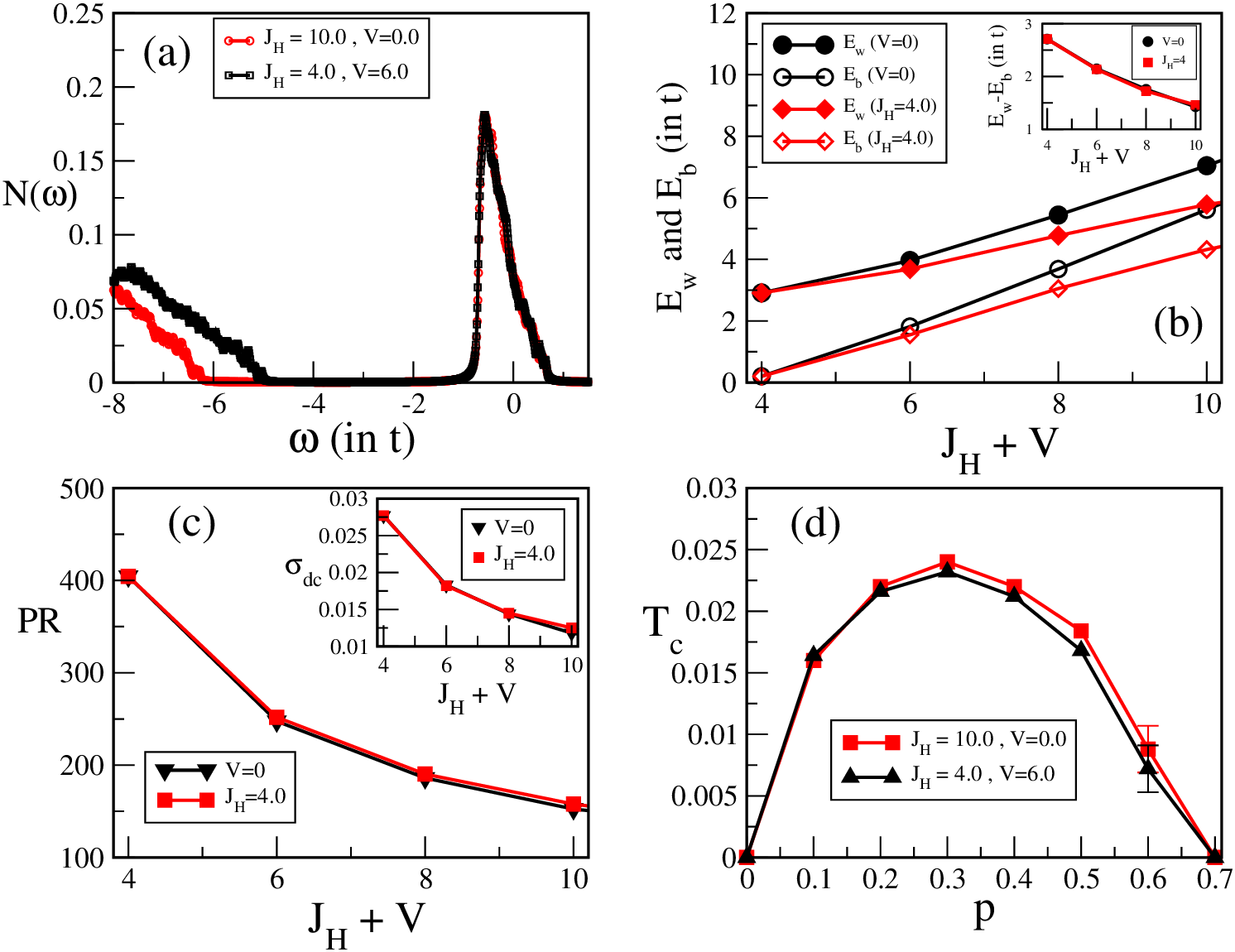}}
\vspace{.2cm}
\caption{\label{fig4}
Presents the comparison of various physical quantities between the $t-J_H$
and the $t-J_H-V$ models at fixed $x$ = 0.25. It compares the (a) density of
states $N(\omega)$ for two sets of parameters ($J_H,V$)=(4,6) and (10,0),
where features of the IBs are shown to match completely.  Fermi energy is
set at zero; (b) variation of the binding energy $E_b$ and the $E_w$ for  
different sets of ($J_H,V$) values. In the x-axis $J_H+V$ is varied in two
different ways: (i) by varying $V$ with fixed $J_H$=4 and (ii) by varying
$J_H$ with fixed $V$=0. Second one is the $t-J_H$ model for which the
physical quantities are replotted from Fig~\ref{fig1}. This shows
that although $E_b$ and $E_w$ differ from one representation to other but
the width of the IBs match well in the whole parameter range [see the inset]; 
(c) variation of the participation ratio and the {\it dc} conductivity
(in units of ${\pi  e^2 }/{\hbar a}$) [in the inset] with ($J_H+V$) values
as in (b). There is one-to-one correspondence between them, and
(d) ferromagnetic windows $T_C$ vs $p$ showing a good match for the two sets of 
parameters as in (a). 
In (a)-(c) the calculations are carried out at fixed $p$ = 0.2 and $T$ = 0.05.
}
\end{figure}

\section{Correspondence between $t-J_H$ and $t-J_H-V$ Models }

The properties of the IB and hence the ferromagnetic window can be tuned
by varying the binding energy of the carriers. Hence it is worthful to
highlight the $t-J_H-V$ model at this point before proceeding with the NNN
hopping term in the Hamiltonian. Here the potential term is represented by
$\sum_m V_m n_m$ with $V_{m}$=V at impurity sites and 0 otherwise. Apart from
the magnetic nature of the Hund's term both $J_H$ and $V$ act as the trapping
centers for the carriers at the impurity sites. So it would be interesting to
check whether the $t-J_H-V$ model can be {\it qualitatively} replaced by a only
$t-J_H$ model or not, in the parameter regime we consider. We benchmark our
results by comparing these two models. Fig.~\ref{fig4}(a) shows the
DOSs for ($J_H, V$) = (10, 0) and (4, 6), where the IB is seen to be unaffected.
Fig.~\ref{fig4}(b) presents the binding energy $E_b$ and the $E_w$ for
different sets of ($J_H,V$) values. In the x-axis $J_H+V$ is defined in
two different ways; (i) by varying V with fixed $J_H$ = 4 representing the
$t-V-J_H$ model and (ii) by varying $J_H$ with fixed V = 0 representing the
$t-J_H$ model. Although the $E_w$ and the $E_b$ differ from one representation
to other for the whole parameter range the widths of the IBs match well
[see the inset]. Consequently, the PR and the conductivity results (see
Fig.~\ref{fig4}(c) and it's inset) for $t-J_H-V$ model is more or less same
to $t-J_H$ model. The comparison of the ferromagnetic windows for both the
set of parameters ($J_H, V$)= (10, 0) and (4, 6) indicate that the $t-J_H-V$ model
can be qualitatively replaced with a suitable choice of $t-J_H$ only, shown in
Fig.~\ref{fig4}(d). Therefore for simplicity we specifically explore the $t-J_H$
model for our further investigations.

\section{Effects of next nearest neighbor hopping for $J_H$=4}

In the recent past Dobrowolska et al.[\hspace*{-3px}\citenum{dobrowolska}] 
demonstrated the existence of a preformed IB in GaMnAs and the 
$T_C$ is decided by the location of the Fermi energy within the impurity 
band. In this picture the states at the center of the impurity 
band are extended resulting in maximum $T_C$. And, the $T_C$ 
gets reduced towards both the top and the bottom ends of the band due to 
localized states. In the process insulator-metal-insulator (I-M-I) transition 
is observed with carrier density. Most importantly, they observed the 
ferromagnetic state in a wide range of hole density $p$ $\sim$ 0.1-0.9. In 
Fig.~\ref{fig2}(c) our FM window ranges only from $p$ = 0 to 0.3 for 
$J_H$ = 4. So now we are going to investigate this mismatch by taking the 
impact of NNN hopping on the carrier mobility and magnetic properties into 
account.

We start with the comparison of the spin-resolved density of states for
$t'$ = 0 and 0.2 at fixed $p$ = 0.2 and T  = 0.004 [see Fig.~\ref{fig5}(a)] 
for which ground states are ferromagnetic. In both the cases the impurity 
band is spin polarized, while the VB remains more or less unpolarized. In
our {\it hole} picture positive $t'$ acts as a localizing agent which can be
visualized from the DOS, where the IB becomes narrow and shifts away from
the VB. This is also apparently clear from the PR shown in
Fig.~\ref{fig5}(b), where the quasiparticle states in case of $t'$=0.2 are
localized compared to $t'$=0 in the whole range of p. It is
also important to note that $t'$ doesn't alter the value of $E_w$ ($\sim$ 3$t$)
which is well within the bandgap of the host. Alternatively, higher $J_H$
can also localize the carriers (as shown in Fig.~\ref{fig1}(a)) and
ultimately boarden the FM window (see Fig~\ref{fig2}(c)), but $E_w$ becomes
larger [see Fig.~\ref{fig1}(b)] than the bandgap which is not physically
acceptable for narrow bandgap host like GaAs.

\begin{figure}[t]
\centerline{
\includegraphics[width=9.0cm,height=7.7cm,angle=0,clip=true]{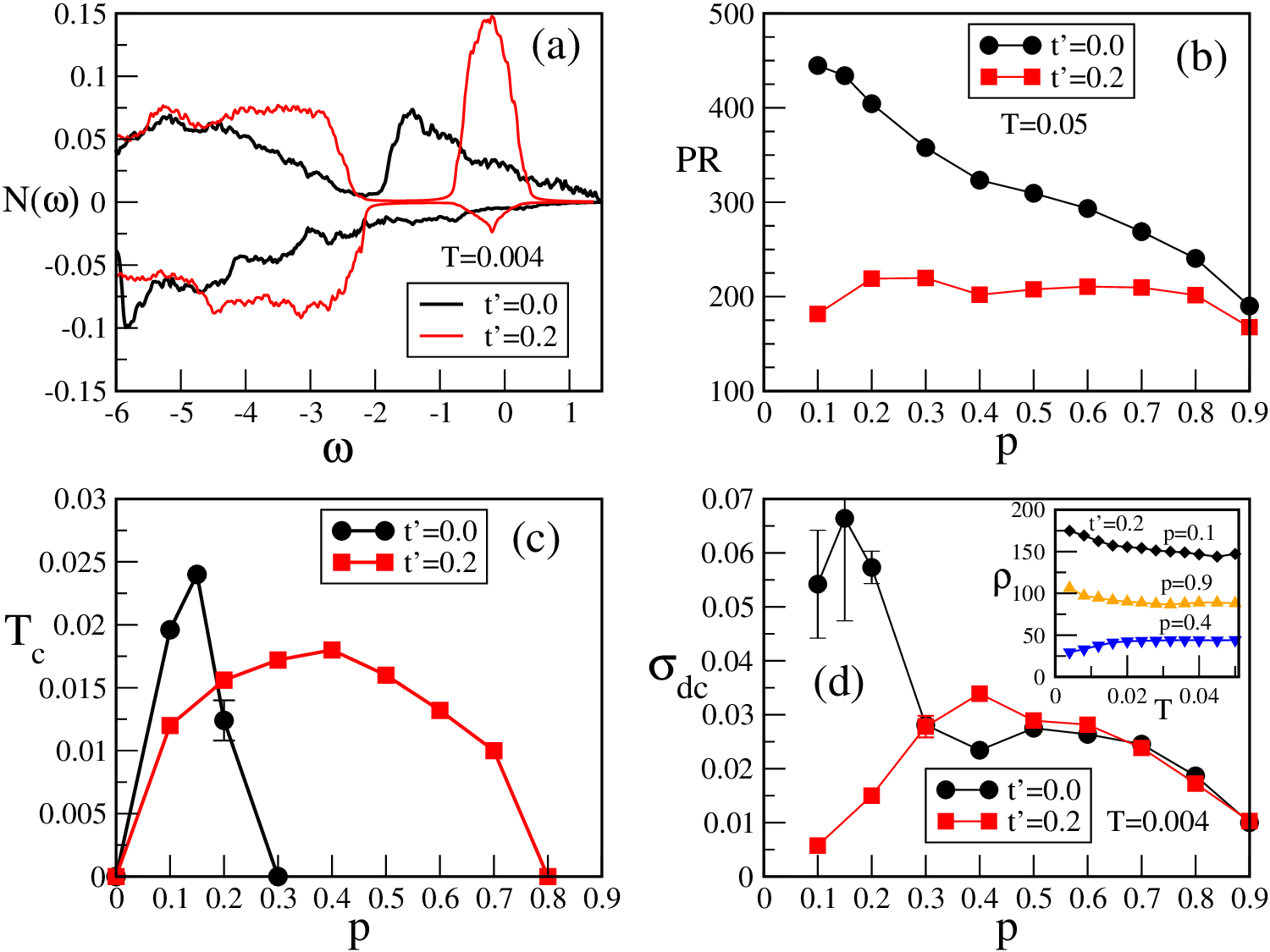}}
\vspace{.2cm}
\caption{\label{fig5}
The effects of the next-nearest-neighbor hopping ($t'$ = +0.2) and it's
comparison to $t'$ = 0 are shown for various physical quantities at fixed
$J_H$ = 4 and $x$ = 0.25. It presents the (a) spin-resolved density of states
at fixed $T$ = 0.004, where the IB shrinks and moves away from the VB due
to carrier localization. The Fermi energy is set at zero; (b) change
in the participation ratio (PR) with $p$ at fixed $T$ = 0.05 showing the
higher degree of localization for $t'$ = +0.2; (c) displays the $t'$-induced
broadening of the ferromagnetic window $T_C$ vs $p$, and 
(d) {\it dc} conductivities (in units of ${\pi  e^2 }/{\hbar a}$) with $p$ at
fixed $T$ = 0.004. I-M-I is confirmed from the 
resistivity (in units of ${\hbar a}/{\pi  e^2 }$) Vs temperature 
plot for different carrier densities, in the inset.
The localization driven I-M-I transition is consistent
with the results presented in (c).
}
\end{figure}

We present the ferromagnetic window, $T_C$ vs $p$, for GaMnAs in
Fig.~\ref{fig5}(c). The $T_C$ optimizes around $p$=0.15 and the ferromagnetism
is restricted to a small window of $p$ = 0-0.3 for $t'$=0. At higher hole
concentration the carrier mobility is suppressed due to larger delocalization
length in GaMnAs, see Fig.~\ref{fig5}(b). One can remobilize the carriers by
reducing their overlap with a mild localization of the carriers which is
stimulated by the NNN hopping parameter $t'$ = 0.2 as shown in Fig.~\ref{fig5}(b).
Consequently, the ferromagnetism is activated and the window
[see Fig.~\ref{fig5}(c)] becomes wider ($p$ = 0-0.8) as observed in the
experiments ($p$ $\sim$ 0.1-0.9)~[\hspace*{-3px}\citenum{dobrowolska,cho}].
In order to correlate the magnetic and transport properties we plot the
low temperature ($T$ = 0.004) {\it dc} conductivity in Fig.~\ref{fig5}(d).
In a carrier-mediated magnetic system a minimum amount of carrier is necessary 
to initiate the magnetic interactions, and at higher $p$ the magnetism is
suppressed due to the decrease in carrier mobility. Hence in these regimes
the system is insulating and in intermediate $p$ the system is metallic resulting
in higher $T_C$. For both $t'$ = 0 and 0.2 conductivity goes through
IMI transition with optimization around the same value of
$p$ as in case of $T_C$ vs $p$ window, which supports the above carrier
localization picture and also qualitatively matches with the experiment~\cite{dobrowolska}.
Resistivity vs Temperature plot in the inset of Fig.~\ref{fig5}(d) explicitly
shows the IMI transition as we increase the hole
density. Experimental data together with our findings hint the presence of 
NNN-hopping in GaMnAs like systems. But further probe and investigations 
are necessary to establish this scenario.

\begin{figure}[t]
\centerline{
\includegraphics[width=9.0cm,height=7.7cm,angle=0,clip=true]{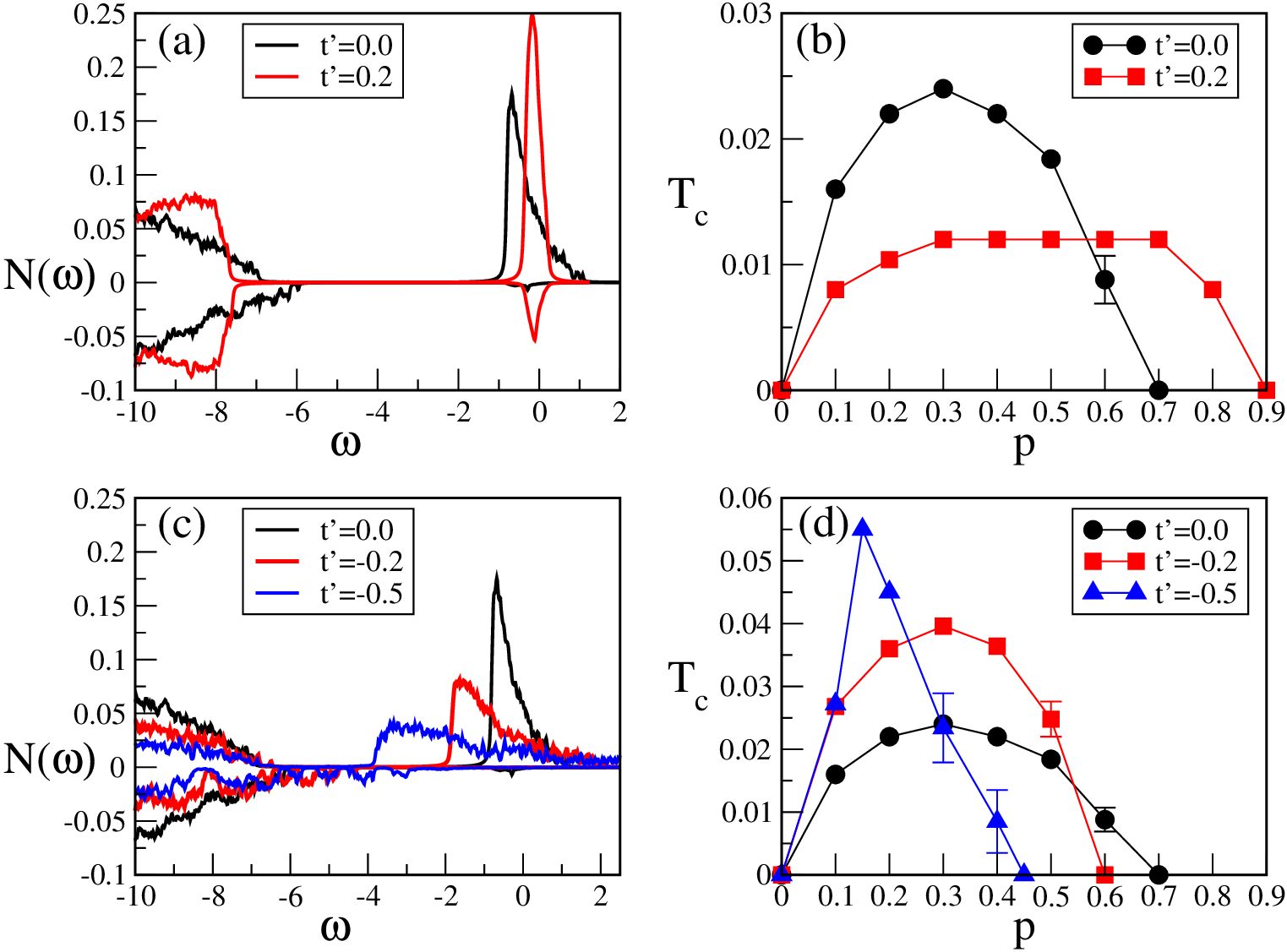}}
\vspace{.2cm}
\caption{\label{fig6}
The effects of the $t'$ = +0.2 and it's comparison to
$t'$ = 0 are presented for (a) the spin-resolved density of states at fixed
$T$ = 0.004, where the IB moves away from the VB due to carrier localization
and (b) the ferromagnetic window $T_C$ vs $p$ showing the localization-induced
broadening. The effects of the $t'$ =-0.2 and -0.5 
with it's comparison to $t'$ = 0 are presented for (c) the spin-resolved
density of states at fixed $T$ = 0.004, where the IB extended towards the
VB due to carrier delocalization, and (d) the ferromagnetic windows showing
the delocalization-induced shrinkening. The Fermi energy is set at zero.
We fixed $J_H$ = 10 and $x=0.25$ for all calculations.
}
\end{figure}

\begin{figure}[t]
\centerline{
\includegraphics[width=9.0cm,height=7.7cm,angle=0,clip=true]{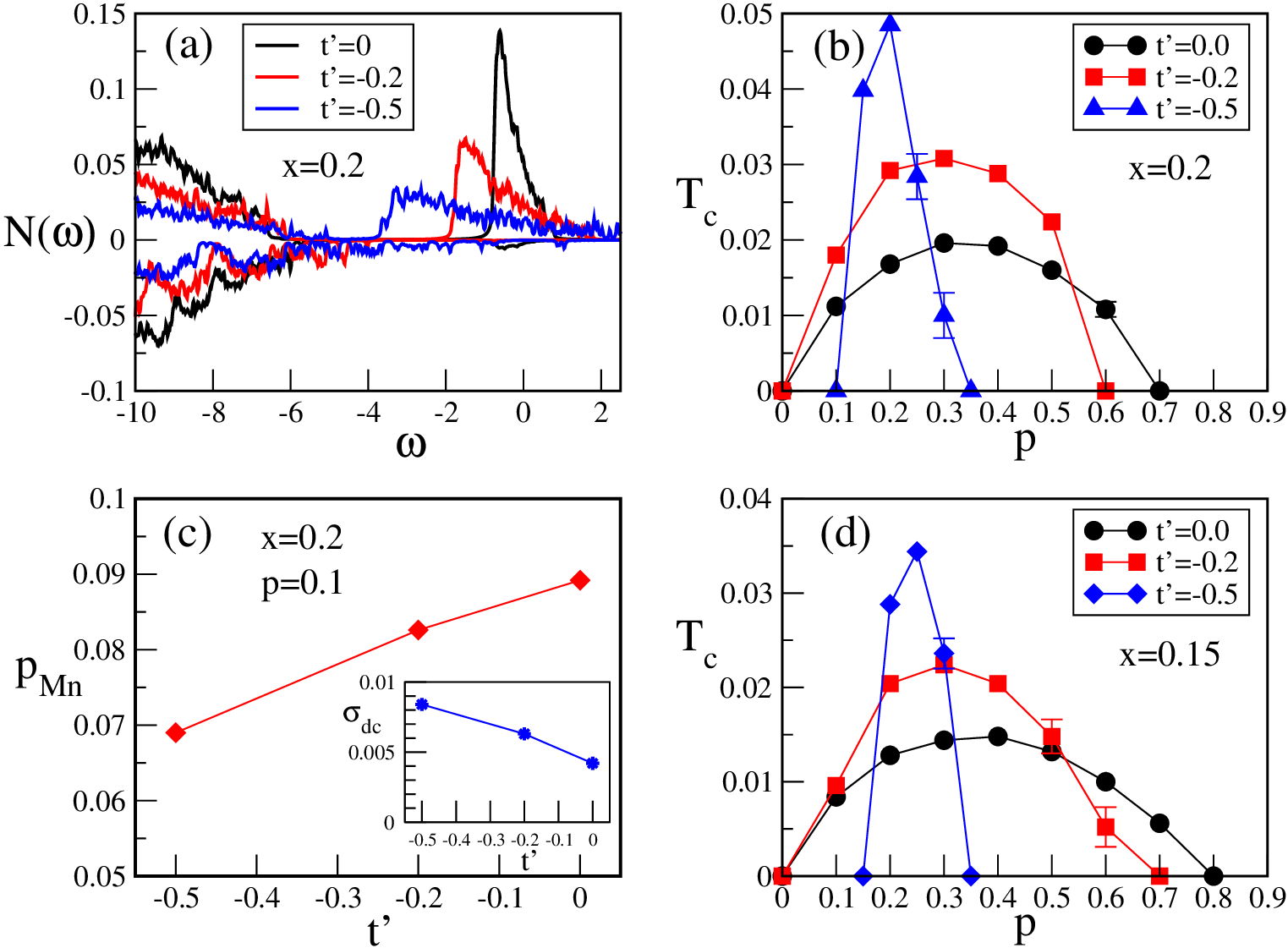}}
\vspace{.2cm}
\caption{\label{fig7}
The effects of the $t'$ = -0.2 and -0.5 with it's
comparison to $t'$ = 0 at fixed $x$ = 0.2 are shown for (a) the spin-resolved
density of states at fixed $T$ = 0.004 and (b) the ferromagnetic windows $T_C$ vs $p$. 
For higher degree of delocalization the FM window becomes significantly narrow. 
(c) plots the average carrier density per magnetic impurity site ($p_{Mn}$) vs $t'$
at fixed $p$=0.1 and $T$ = 0.05, which reveals the out flow of carriers to
non-magnetic sites with delocalization. In consequence the overall conductivity
of the system increases as shown in the inset. We find similar narrowing of FM
window for $x$ = 0.15 and presented in (d). We considered $J_H$ = 10 for all
calculations.
}
\end{figure}

\section{Effects of next nearest neighbor hopping for $J_H$=10}

Now we study the GaMnN with the chosen Hund's coupling $J_H$ = 10. The
spin-resolved DOS and the FM windows are evaluated with the same set of
parameter values as in Fig.~\ref{fig5}. Fig.~\ref{fig6}(a) and (b) show that
the effect of NNN hopping on the IB and FM window are qualitatively similar
as in the case of $J_H$ = 4. Quantitatively, the effect of localization
due to $t'$ is much more prominent for $J_H$ = 10 and as a result the deduction
of $T_C$ is significant. But, the electronic structure calculations reveal that
the Ga defects in GaMnN introduce states between the VB and the IB which
depopulate the IB and in turn destroy the ferromagnetism in GaMnN~\cite{mahadevan}.
We mimic the situation by introducing negative NNN hopping which delocalize the
carriers and consequently broaden the IB towards the VB. This can be seen
from the DOS plotted in Fig.~\ref{fig6}(c) for $t'$ = -0.2 and -0.5 along
with $t'$ = 0 at fixed $p$ = 0.2 and $T$ = 0.004. Note that the binding
energy $E_b$ decreases but $E_w$ remains more or less unaffected
(i.e. $E_w$ is within the bandgap of host GaMnN). As positive and negative $t'$
play opposite roles in the system, so the ferromagnetic window shrinks
and the optimum $T_C$ increases with carrier delocalization, as shown
in Fig.~\ref{fig6}(d).

The solubility of Mn in GaAs and GaN is low, so we establish our findings
by calculating the ferromagnetic windows for lower impurity concentrations. 
First we consider $x$ = 0.2 and the results for the spin-resolved DOS and
the FM windows are presented in Fig.~\ref{fig7}(a) and (b) respectively. 
The IBs show qualitatively similar evolution with $t'$ as in case of
$x$ = 0.25. Apart from the disappearance of ferromagnetism in the higher
$p$ regime, interestingly, the magnetism also vanishes for very low carrier
densities for $t'$= -0.5 making the FM window furthermore narrow. Note
that we have considered the relative carrier density i.e. number of carrier
per Mn impurity site as in experiments. So, in the low $x$ and lower $p$
regime the magnetic sites accumulate a tiny amount of holes resulting in a
weaker magnetic interactions. Here, if we increase the carrier mobility the
holes get deplete from the magnetic to the non-magnetic sites which further
suppresses the spin-spin couplings. The out flow of carriers is displayed in
Fig.~\ref{fig7}(c) where we plot the average carrier density at the
magnetic sites $p_{Mn}$ vs $t'$ at fixed $p$=0.1. Eventually, the
ferromagnetism vanishes at lower $p$ as in case of $t'$ = -0.5. On the
other hand, the overall conductivity of the system increases with the
degree of delocalization as shown in the inset. We find similar results
for $x$ = 0.15 [Fig.~\ref{fig7}(d)]. The vanishing ferromagnetism in
both lower and higher $p$ regimes for $t'$ = -0.5 makes the ferromagnetic
window significantly narrow, which suppresses the probability of getting a
FM state. In experiments the presence of defects makes the sample
preparation very crucial and our results indicate that unless the system
has a favourable combination of $x$ and $p$ in a narrow window then
there is a higher chance to observe either low $T_C$ or absence of
ferromagnetism. In addition, the sharp increase in the optimum $T_C$
in a thin window of $p$ clarifies the room temperature ferromagnetism
occasionally observed in experiments.

\section{Conclusions}

In conclusion, we investigated the magnetic and the transport properties
of III-V DMSs using a classical Monte-Carlo method within the $t-t'$ Kondo
lattice model on a simple cubic lattice. We have shown that the carrier
mobility induced by the NNN hopping $t'$ plays a vital role in determining
the ferromagnetic states in both GaMnAs and GaMnN like systems. In case of
GaMnAs a small positive $t'$ (that helps to localize the carriers) 
is shown to be necessary to
reproduce the robustness of the ferromagnetic states in a wide range of
carrier concentration as observed in experiments. On the other hand, if
we delocalize the carriers by activating negative $t$' the ferromagnetic window
significantly shrinks with an enhancement of the optimum value of $T_C$ in
GaMnN. We correlate our findings with the experimental results and suggest
that Ga like vacancy in GaMnN that depopulate the IB triggers high $T_C$
in low hole density. In reality, the presence of intrinsic defects is
inevitable and also the carrier density is not controllable. So the
probability of having an optimal amount of holes in a narrow regime in
Ga defected GaMnN is very low. This could be the reason of occasional
appearance of ferromagnetism and in turn keeps the high $T_C$ issue of
GaMnN unresolved till date.

\noindent \\
Acknowledgment: We acknowledge use of Meghnad2019 computer cluster at SINP.

\end{document}